\documentclass[12pt]{JHEP3}
\usepackage{epsf}
\newcommand{\be}{\begin{eqnarray}}
\newcommand{\ee}{\end{eqnarray}}

\newcommand{\com}[2]{\left[#1,#2\right]}

\author{Rikard von Unge\\
Institute for Theoretical Physics and Astrophysics\\
Faculty of Science, Masaryk University\\
Kotl\'{a}\v{r}sk\'{a} 2, CZ-611 37, Brno, Czech Republic\\
\email{unge@monoceros.physics.muni.cz}}

\abstract{We extend the path-integral formalism
 for Poisson-Lie T-duality to include the case of Drinfeld
doubles which can be decomposed into bi-algebras in more than one
way. We give the correct shift of the dilaton, correcting a mistake in
the litterature. We then use the fact that the six dimensional Drinfeld
doubles have been classified to write down all possible conformal Poisson-Lie
T-duals of three dimensional space times and we explicitly work out
two duals to the constant dilaton and zero anti-symmetric tensor Bianchi
type V space time and show that they satisfy the string equations
of motion. This space-time was previously thought to have no duals
because of the tracefulness of the structure constants.
}
\title{Poisson-Lie T-plurality}

\begin{document}
\section{Introduction}
The importance of target space (or T for short) duality in string
theory \cite{KY,SS} can hardly be overstated. It has played a
fundamental role in gaining non-perturbative information about string
theory. In its original appearance it can be understood as a symmetry
of the string background equations of motion on a background which has
abelian isometries \cite{busch1}. Given such a background there exists
a well defined procedure in the path-integral setting for finding the
dual background \cite{mandv}. Carefully regularizing the
determinants that 
appear when performing the path-integral this procedure also leads to
a shift in the dilaton \cite{busch2} which is necessary for the dual
background to be consistent with string theory, i.e. for the dual
background to satisfy the string beta functions. Later, this procedure
was generalized to the case where the isometries of the background are
non-abelian \cite{nab}. In the non-abelian case there appeared two new
features which are not present in the original abelian duality
setting. The first is, as was discovered in \cite{venez}, that for
some backgrounds, even including the dilaton shift, the dual is not
conformal (it does not satisfy the beta functions). This was later
explained \cite{manda,aagl} in terms of a mixed gravitational/gauge
anomaly which appears when the structure constants of the initial
isometry are not traceless ($f_{ab}^a\neq0$). The second puzzling
feature was that the dual space-time one ends up with following the
standard non-abelian T-dualization procedure in general does not have
isometries making it impossible to perform the duality once more to
get back to what one started with, thus making the name ``duality''
somewhat of a misnomer. This second feature was understood in
\cite{ks} (see also \cite{ck}) where Poisson-Lie T-duality was
introduced. In that paper it was explained how abelian and
non-abelian T-duality can be embedded in a larger algebraic framework
where the basic object is a so called Drinfeld double (more on this in
the next section). They showed that the sigma models one ends up with
are dual to each other by showing that there exists canonical
transformations between them. In particular, there is no need for a
sigma model to have isometries for the duality to work. However, in
this approach it is not clear how the dilaton should transform
since the dilaton transformation is a quantum effect. A first look at
quantum effects in the context of Poisson-Lie T-duality was 
made in \cite{quant1,us}. In \cite{us} a path-integral approach for
Poisson-Lie T-duality was introduced (see also \cite{related}). It was
found that just as in 
non-abelian T-duality \cite{manda,aagl}, a group with traceful
structure constants introduces an anomaly which spoils the conformal
invariance. Also, the way the dilaton should transform in order to
preserve conformal invariance was found. Unfortunately, due to the
lack of explicit examples of Drinfeld doubles at that time it was not
possible to test this prescription on a concrete example. Therefore a mistake
in the dilaton transformation formula was not discovered. Further
quantum properties were studied and discussed in
\cite{quant3,quant4,quant5,quant6,quant7}. By now Poisson-Lie
T-duality is quite well studied and generalizations have been made in
several directions. It has been studied in the context of
supersymmetry \cite{susy1,susy2,susy3,susy4,susy5,susy6,susy7,susy8},
in the context of open strings and D-branes \cite{db1,db2} and in the
context of mirror symmetry \cite{mirror1,quant5}.

Recently all six dimensional Drinfeld doubles have been classified
\cite{libor,class2}. This provides a large class of examples on which
Poisson-Lie T-duality can be tested. Furthermore, in \cite{libor} it
was explicitly shown how most Drinfeld doubles actually can be
decomposed into bi-algebras in more than one way\footnote{This was
anticipated already in \cite{ks}}. This makes the
duality much richer - one should maybe speak about plurality
instead. Quite surprising is that ordinary non-abelian T-duality where
the isometry {\em does} have traceful structure constants and thus
will be anomalous can be embedded in Poisson-Lie T-duality in such a
way that if the Drinfeld double admits more than one decomposition and
in that decomposition at least one of the algebras have trace free
structure constants, it is possible to find a dual {\em and} conformal
background (or backgrounds depending on how many such decompositions
there are) to the initial background. The purpose of this paper is to
generalize the path-integral formulation of Poisson-Lie T-duality
given in \cite{us} to include the case of Drinfeld doubles
decomposable in more than one way. We give the correct dilaton and
background transformations and use the Bianchi V space time of
\cite{venez} to show the power of Poisson-Lie T-duality. This
space-time was used in \cite{venez} to show that non-abelian T-duality
does not work when the structure constants are traceful. Here we show
how Poisson-Lie T-duality gives us two new space-times which we give
explicitly and check that they indeed are solutions to the string
equations of motion.

The paper is organized as follows: In section \ref{dd} we give
definitions of and introduction to Drinfeld doubles and Poisson-Lie
T-duality. In section \ref{general} we give the general formalism and
derive the general formulas for the background and dilaton shift. In
section \ref{all} we give a general discussion of all (possibly) {\em
conformal} Poisson-Lie T-duality chains in three dimensions. In
section \ref{explicit} we calculate explicitly the duals of the
Bianchi V space time used in \cite{venez} and in section \ref{disc} we
conclude and discuss open questions.

\section{Drinfeld doubles and T-duality}\label{dd}
A Drinfeld double $D$ is defined as a connected Lie group such that its
Lie algebra {\cal D}, equipped with a symmetric ad-invariant bilinear form
$\langle\bullet,\bullet\rangle$, can be decomposed into a pair of
sub-algebras ${\cal G}$ and $\tilde{\cal G}$, maximally isotropic with
respect to $\langle\bullet,\bullet\rangle$ and such that the algebra
of the Drinfeld double is the direct sum of the two sub-algebras.

The dimensions of the sub-algebras have to be equal and one can choose
a basis in each of the sub-algebras $T_{a}\in {\cal G}$,
$\tilde{T}^{a}\in \tilde{\cal G}$ such that
\be
\langle T_a , T_b \rangle = \langle \tilde{T}^a, \tilde{T}^b \rangle =
0,\nonumber\\
\langle T_a , \tilde{T}^b\rangle = \delta_{a}^{b}.
\ee
Assuming that the two sub-algebras have the form
\be
\left[T_a , T_b\right] &=& f_{ab}^{\;\;c}T_c,\nonumber\\
\left[\tilde{T}^a , \tilde{T}^b\right] &=& 
\tilde{f}^{ab}_{\;\;c} \tilde{T}^c.
\ee
one can use the ad-invariance of $\langle \bullet,\bullet\rangle$ to
show that the remaining commutation relations must be
\be
\left[T_a,\tilde{T}^b\right] = f_{ca}^{\;\;b}\tilde{T}^c
+ \tilde{f}^{bc}_{\;\;a} T_c.
\ee
Furthermore, the Jacobi identities on the Drinfeld double implies the
relation between the structure constants
\be\label{jacobi}
f_{ab}^{\;\;e}\tilde{f}^{cd}_{\;\;e} =
f_{ae}^{\;\;c}\tilde{f}^{ed}_{\;\;b}
+f_{ae}^{\;\;d}\tilde{f}^{ce}_{\;\;b}
-f_{be}^{\;\;c}\tilde{f}^{ed}_{\;\;a}
-f_{be}^{\;\;d}\tilde{f}^{ce}_{\;\;a}.
\ee

In the seminal paper \cite{ks} it was shown that one can use this
algebraic structure to extend the notion of a background with
isometries to backgrounds with so called generalized
isometries. Starting with a sigma model action defined on a Lie group
\be
 \int \left(g^{-1}\partial g \right)^{a} E_{ab}
   \left(g^{-1}\bar{\partial}g\right)^{b} = 
 \int \partial x^{i} F_{ij} \bar{\partial}x^{j},
\ee
where we have used the left invariant frames to write the model in
coordinates $F_{ij} = e^{a}_i E_{ab} e^b_j$, we say that the
background $F_{ik}$ has
(non-abelian) isometries if ${\cal L}_{v_a} F_{ik} = 0$ for some
vector field $v_a$ satisfying $\left[v_a ,v_b\right] = f_{ab}^{c}v_c$
However, if the background instead satisfies
satisfies
\be
{\cal L}_{v_c} F_{ik} = F_{ij}v^{j}_a \tilde{f}^{ab}_{c} v^l_b F_{lk},
\ee
for some constants $\tilde{f}^{ab}_c$, we say that the background has
{\em generalized} isometries. The consistency condition on
the lie derivative $\left[{\cal L}_{v_a},{\cal L}_{v_b}\right] = {\cal
L}_{\left[v_a,v_b\right]}$ then implies the relation (\ref{jacobi})
showing that this construction leads naturally to the Drinfeld
double. In \cite{ks} it was shown that it
is possible to define an equivalent but dual sigma model with a background
$\tilde{F}$ satisfying
\be
{\cal L}_{\tilde{v}^c} \tilde{F}^{ik} = 
\tilde{F}^{il}v^a_l f_{ab}^c v^b_n \tilde{F}^{mk},
\ee
where $\tilde{F}$ is related to $F$ at the ``origin'' of the Drinfeld
double by the relation $\tilde{F}(\tilde{x}=0) = F^{-1}(x=0)$ where
$x$ are the coordinates on the original space-time and $\tilde{x}$ are
the coordinates on the dual.

\section{General formalism}\label{general}
The path integral formulation of Poisson-Lie T-duality was given in
\cite{us}. The parent action taken as the starting point of the
duality is a WZW-model\footnote{Notice that this does not mean that
the final sigma models are of WZW type. They are simply sigma models
defined on a Lie group.} defined on the Drinfeld double but with an
extra, chiral constraint. In formulas we have
\be
{\cal Z} = \sqrt{\det G^{(0)}}\int {\cal D}l \;
\delta\left[ \langle l^{-1}\partial l,\tilde{T}^a\rangle E^{(0)}_{ab}
-\langle l^{-1}\partial l , T_b \rangle\right]
e^{-I\left[ l \right]},
\ee
where $E^{(0)}_{ab}$ is a constant (or possibly dependent on the
spectator coordinates\footnote{Spectator coordinates are coordinates
that do not explicitly take part in the dualization. For instance, if
we discuss 6 dimensional Drinfeld doubles, the duality will be between
3 dimensional spaces which means that for the bosonic string there
would be 23 spectator coordinates.}) matrix, $G^{(0)}$ is the symmetric
part of $E^{(0)}$ and 
\be
I\left[ l\right] = \frac{1}{4\pi}\int \langle l^{-1}\partial l ,
l^{-1}\bar{\partial}l\rangle + \frac{1}{12\pi} \int
\frac{1}{d}\langle l^{-1}dl,\left[l^{-1}dl,l^{-1}dl\right]\rangle.
\ee
Notice that the generators $(T,\tilde{T})$ defines a ``canonical''
decomposition of the double. To work with arbitrary decompositions
we define the transformation matrix
\be
\left(\begin{array}{c}
T\\ \tilde{T} \end{array}\right) =
\left(\begin{array}{cc}
F & G\\ H & K \end{array}\right)
\left(\begin{array}{c}
U\\ \tilde{U} \end{array}\right).
\ee
The group associated with the generators $U$ will be the group over
which the final sigma model is defined and the group associated with
the generators $\tilde{U}$ will be the auxiliary group which is
integrated out. This formalism includes the canonical decomposition by
choosing the decomposition matrix as
\be\label{trivial}
\left(\begin{array}{cc}
F & G\\ H & K \end{array}\right) =
\left(\begin{array}{cc}
1& 0\\0& 1 \end{array}\right),
\ee
and the canonical dual by choosing the decomposition matrix as
\be\label{basic}
\left(\begin{array}{cc}
F & G\\ H & K \end{array}\right) =
\left(\begin{array}{cc}
0& 1\\1& 0 \end{array}\right).
\ee
Defining the matrices
\be\label{matrices}
\mu^{ab}(g) &=& \langle g \tilde{U}^{a} g^{-1}, \tilde{U}^{b}\rangle,
\nonumber\\
\nu_{a}^{\;\;\;b}(g) &=& \langle U_{a} , g\tilde{U}^{b} g^{-1}
\rangle,\nonumber\\
M &=& K^{t}E^{(0)} - G^{t},\\
N &=& F^{t} - H^{t} E^{(0)},\nonumber
\ee
where $g$ is a group element in the group generated by the $U$
generators, and decomposing the general group element $l =
\tilde{g}g$, with $\tilde{g}$ in the group generated by the
$\tilde{U}$ generators, the following happens:
\begin{itemize}
\item The measure ${\cal D}l$ being the left invariant Haar measure on
the Drinfeld double, splits into
\be
{\cal D}l = {\cal D}\tilde{g}{\cal D}g \det\nu^{-1}(g).
\ee
\item The delta function can be written as
\be
\delta\left[ J M - \tilde{A}\left(\nu^{-1}\right)^{t}
\left(N+\mu\nu M\right)\right],
\ee
where $J^{a} = \langle \tilde{U}^{a},g^{-1}\partial g\rangle$ and
$\tilde{A}_{a} = \langle U_{a}, \tilde{g}^{-1}\partial
\tilde{g}\rangle$.
\item The action, using the Polyakov-Wiegmann identity \cite{PW} and the fact
that $I\left[g\right] = I\left[\tilde{g}\right] = 0$ which follows from
the skew symmetry of $\langle \bullet , \bullet \rangle$, becomes
$I\left[\tilde{g}g\right] =  \frac{1}{4\pi}\int
\tilde{A}\left(\nu^{-1}\right)^{t} \bar{J}$.
\end{itemize}
The integral over the variable $\tilde{g}$ can be performed by
changing variables to $\tilde{A} = \tilde{g}^{-1}\partial\tilde{g}$
but doing so we pick up a non-trivial determinant
\be
 {\cal D} \tilde{g} = {\cal D} \tilde{A} 
 \frac{1}{\det\left( \partial + \left[ \tilde{A},\cdot\right]\right)},
\ee
which, if the structure constants of the algebra generated by $\tilde{U}$
has non-zero trace (i.e. $\tilde{f}_{ab}^{\;\;\;\;b} \neq 0$), has gauge and
gravitational anomalies ruining the T-duality as was observed
in \cite{manda,aagl}. In the following we will assume that the
auxiliary group is always anomaly free so that we do not have to
worry about these factors.

Assuming that we have chosen the
$\tilde{U}$ algebra to be anomaly free, we can go on and integrate out
the auxiliary group giving us the sigma model
\be
 \int {\cal D} g \det\nu^{-1} \left[
\det \left(\left(\nu^{t}\right)^{-1}
 \left(N+\mu\nu M\right)\right) \right]_{\rm reg}^{-1}
 e^{-\frac{1}{4\pi}\int J \left(NM^{-1} + \mu\nu\right)^{-1}\bar{J}},
\ee
The determinant in the parenthesis, coming from the functional
integration over $\tilde{A}$ is not well defined and needs to be
regularized. Following \cite{tseyt1,tseyt2} we can write it as
\be
\left[ \det \left(\left(\nu^{t}\right)^{-1}
 \left(N+\mu\nu M\right)\right) \right]_{\rm reg}^{-1} &=&
\det \left(\left(\nu^{t}\right)^{-1}
 \left(N+\mu\nu M\right)\right)^{-1} 
\nonumber\\ &&
e^{\frac{1}{8\pi}\int R^{(2)}
\ln\left( \det \left(\left(\nu^{t}\right)^{-1}
 \left(N+\mu\nu M\right)\right) \right)}.
\ee
The determinant in front of the exponential is formally the same as
the determinant that we started with but it is now regularized and
together with the $\sqrt{\det G^{(0)}}$ and the $\det\nu^{-1}$ in the
original measure it gives us a final path-integral measure
\be
{\cal D}g \frac{\sqrt{\det G^{(0)}} \det E}{\det M}.
\ee
It should be possible to show that this can always be written as
${\cal D}g \sqrt{\det G}$ where $G$ is the symmetric part of $E$ so
that the measure is always the correct coordinate invariant sigma
model measure. We have not done this but instead we have checked it
explicitly in every example we have calculated.
The factor in the exponential can be absorbed in a shift of the
dilaton. In summary this gives us a sigma model with the correct
integration measure and the background, including the dilaton shift
\be
 E &=& \left(N M^{-1} + \mu\nu\right)^{-1},\nonumber\\
 \phi &=& \phi^{(0)} + \ln\det M - \ln\det E - \ln\det \nu.
\ee
One can check that choosing the trivial decomposition matrix
(\ref{trivial}), one gets the background $E = \left(
\left(E^{(0)}\right)^{-1} +\mu\nu \right)^{-1}$, and choosing the basic dual
decomposition matrix (\ref{basic}), one gets the background $E =
\left(E^{(0)} + \mu\nu\right)^{-1}$.\footnote{Notice here that the matrices
$\mu(g)$ and $\nu(g)$ differ in these two examples. They are always
calculated for a particular decomposition since $g$ is defined to be
in the group generated by the $U$ generators. Therefore $\mu$ and
$\nu$ are decomposition dependent} Up to the last term in the dilaton
shift this agrees with the standard formulas as given in
\cite{ks,us}. This term appears when carefully regulating all
determinants and is essential for the dual model to be conformal. That
this term was left out in \cite{us} is presumably the source of the
discrepancy in the dilaton shift of \cite{quant4}.

\section{All conformal three dimensional duals}\label{all}
In \cite{libor} all 6-dimensional Drinfeld doubles were
found. Particular care was taken to identify different decompositions
into bi-algebras. In all cases where the same double can be decomposed
in different ways an explicit transformation matrix, relating the
generators in the two cases, were given.
{}From the classification in \cite{libor} we can read off all possible
{\em conformal} duality chains in three dimensions. Using their notation
$(G|\tilde{G})$\footnote{The notation is such that $G$ and $\tilde{G}$
are numbers referring to its Bianchi type. In the case
where there are more parameters they are indicated by various
indices. For a complete definition see \cite{libor}} and taking care
to always write the auxiliary group 
to be integrated over in the second position, we find
\begin{itemize}
\item $(5_b|9)\leftrightarrow (5_b.ii|8) \leftrightarrow (5_b.ii|7_0)$.
\item $(5_b.i|8) \leftrightarrow (5.iii|6_0)$.
\item $(9|1) \leftrightarrow (1|9)$.
\item $(8|1) \leftrightarrow (1|8) \leftrightarrow (5.iii|8) 
\leftrightarrow (5.i|7_0) \leftrightarrow (5_i|6_0) \leftrightarrow
(5|2.ii)$.
\item $(4|2_{b}.iii) \leftrightarrow (4_b|7_0) \leftrightarrow (4_{-b}.i|6_0)$.
\item $(7_a|1) \leftrightarrow (7_a|2.i) \leftrightarrow (7_a|2.ii)$.
\item $(6_a|1) \leftrightarrow (6_a|2)$.
\item $(5|1) \leftrightarrow (5|2.i) \leftrightarrow (6_0|1)
\leftrightarrow (1|6_0) \leftrightarrow (5.ii|6_0)$.
\item $(4|1) \leftrightarrow (4|2.i) \leftrightarrow (4|2.ii)
\leftrightarrow (6_0|2) \leftrightarrow (2|6_0) \leftrightarrow
(4.ii|6_0)$.
\item $(3|1) \leftrightarrow (3|2)$.
\item $(7_0|1) \leftrightarrow (1|7_0)$.
\item $(7_0|2.i) \leftrightarrow (2.i|7_0)$.
\item $(7_0|2.ii) \leftrightarrow (2.ii|7_0)$.
\item $(2|1) \leftrightarrow (1|2)$.
\item $(2|2.i) \leftrightarrow (2.i|2)$.
\item $(2|2.ii) \leftrightarrow (2.ii|2)$.
\item $(1|1)$.
\end{itemize}
This list should be seen as the three dimensional Poisson-Lie
equivalent of the orbits of the abelian $O(d,d,R)$ duality groups
modulo the automorphisms of each of the non isomorphic decompositions
by itself as was discussed in \cite{ks}. That is, the full duality
group would include the automorphisms of each of the algebras and the
elements which takes one from one decomposition to the other in the way
indicated above. These elements are given exactly by the matrices in
\cite{libor}. It it also possible to have subdualities for Drinfeld
doubles which contains lower dimensional sub Drinfeld doubles. A first
step in trying to find these duality groups was taken in \cite{moduli}.

\section{An explicit example. Duals of Bianchi V}\label{explicit}
As an explicit example we calculate duals of a Bianchi type V
space time with constant dilaton and no antisymmetric tensor field
\be\label{flat}
ds^2 &=& -dt^2 + t^2\left(dx^2 + e^{2x}(dy^2 +dz^2)\right),
\nonumber\\
 b   &=& 0,\\
 \phi &=& 0.\nonumber
\ee
This is a solution to the string beta function equations since
the metric is in fact flat. This is an appropriate metric
to choose for illustration since it was for this metric that it was
first noticed that non-abelian duals of space times with isometry
group with non traceless structure constants are not conformal
\cite{venez}. Later it was discovered that this was because of the
mixed gravitational/gauge anomaly coming from the jacobians in the
path-integral measure. By taking care and always choosing the
auxiliary group to be anomaly free, we are guaranteed to get only
conformal dual models.

\bigskip
\begin{flushleft}
\underline{$(5|1)$:}
\end{flushleft}
\bigskip
We take $(5|1)$ as the canonical decomposition
The full algebra is given by
\be
\com{T_1}{T_2} = -T_2 & \com{T_2}{T_3} = 0 & \com{T_3}{T_1} = T_3\nonumber\\
\com{\tilde{T}^1}{\tilde{T}^2} = 0 & 
\com{\tilde{T}^2}{\tilde{T}^3}=0 & \com{\tilde{T}^3}{\tilde{T}^1} =
0\nonumber\\
\com{T_1}{\tilde{T}^1} = 0& \com{T_2}{\tilde{T}^1} = 0& 
\com{T_3}{\tilde{T}^1} = 0\\
\com{T_1}{\tilde{T}^2} = \tilde{T}^2& \com{T_2}{\tilde{T}^2} = -\tilde{T}^1& 
\com{T_3}{\tilde{T}^2} = 0\nonumber\\
\com{T_1}{\tilde{T}^3} = \tilde{T}^3& \com{T_2}{\tilde{T}^3} = 0&
 \com{T_3}{\tilde{T}^3} = -\tilde{T}^1\nonumber
\ee
The parametrization of a general group element we choose as
\be
 l = \tilde{g}g = e^{\tilde{x}_3\tilde{T}^3}
 e^{\tilde{x}_2\tilde{T}^2}e^{\tilde{x}_1\tilde{T}^1}
     e^{x_3 T_3} e^{x_2 T_2} e^{x_1 T_1},
\ee
giving
\be\label{liformV}
g^{-1}dg = dx_1 T_1 + e^{x_1} dx_2 T_2 + e^{x_1} dx_3 T_3.
\ee
Calculating the matrices (\ref{matrices}) for this decomposition, we get
\be
\mu &=& 0,\nonumber\\
\nu &=& \left(\begin{array}{ccc}
1 & -x_2 e^{x_1} & -x_3 e^{x_1}\\
0 & e^{x_1} & 0\\
0 & 0 & e^{x_1}
\end{array}\right),\nonumber\\
M &=& E^{(0)},\\
N &=& 1, \nonumber
\ee
which gives us an $E=E^{(0)}$ as expected for the canonical
decomposition. For this to give us the background (\ref{flat}) we see
that we have to choose $E^{(0)}= t^2 {\bf 1}$. Since $\mu = 0$ there
is no anti-symmetric tensor. However, since we want the total dilaton
to be zero we need to choose
\be
\phi^{(0)} = \ln\det E + \ln\det\nu - \ln\det M = \ln\det \nu = 2x_1.
\ee
The measure factor can also be calculated. Since $\det G^{(0)} = t^6$
and in this case $\det E = \det M = t^6$ it comes out as
\be
 {\cal D} g \frac{t^3 t^6}{t^6} = {\cal D}g \sqrt{\det G},
\ee
where $G$ is the symmetric part of $E$ so that the full measure is
indeed the correct coordinate invariant measure of the sigma model.

\bigskip
\begin{flushleft}
\underline{$(6_0|1)$:}
\end{flushleft}
\bigskip
{}From \cite{libor} we find the non-trivial decomposition matrix
\be
 \left(\begin{array}{c}
 T\\ \tilde{T}\end{array}\right) =
\left(\begin{array}{cccccc}
 0 & 0 & -1 & 0& 0& 0\\
 0& 0& 0& 1& 1& 0\\
 -1& 1& 0 & 0& 0& 0\\
 0& 0& 0& 0& 0& -1\\
 \frac{1}{2}& \frac{1}{2}& 0 & 0& 0& 0\\
 0& 0& 0& -\frac{1}{2} & \frac{1}{2} & 0
\end{array}\right)
\left(\begin{array}{c}
U\\ \tilde{U}\end{array}\right).
\ee
The algebra is
\be
\com{U_1}{U_2} = 0 & \com{U_2}{U_3} = U_1 & \com{U_3}{U_1} = -U_2\nonumber\\
\com{\tilde{U}^1}{\tilde{U}^2} = 0 & 
\com{\tilde{U}^2}{\tilde{U}^3}=0 & \com{\tilde{U}^3}{\tilde{U}^1} =
0\nonumber\\
\com{U_1}{\tilde{U}^1} = 0& \com{U_2}{\tilde{U}^1} = -\tilde{U}^3& 
\com{U_3}{\tilde{U}^1} = \tilde{U}^2\\
\com{U_1}{\tilde{U}^2} = -\tilde{U}^3& \com{U_2}{\tilde{U}^2} = 0& 
\com{U_3}{\tilde{U}^2} = \tilde{U}^1\nonumber\\
\com{U_1}{\tilde{U}^3} = 0& \com{U_2}{\tilde{U}^3} = 0&
 \com{U_3}{\tilde{U}^3} = 0\nonumber
\ee
Decomposing an arbitrary group element as
\be
 l = \tilde{g}g = e^{\tilde{y}_1\tilde{U}^1}
 e^{\tilde{y}_2\tilde{U}^2}e^{\tilde{y}_3\tilde{U}^3}
     e^{y_1 U_1} e^{y_2 U_2} e^{y_3 U_3},
\ee
which is equivalent to a change of coordinates
\be\label{cchange}
x_1 &=& -y_3,\nonumber\\
x_2 &=& \frac{1}{2}\left(\tilde{y}_1+\tilde{y}_2\right),\nonumber\\
x_3 &=& \frac{1}{2}\left(y_2- y_1\right),\nonumber\\
\tilde{x}_1 &=& -\tilde{y}_3-\frac{1}{2}\left(y_1+y_2\right)
\left(\tilde{y}_1+\tilde{y}_2\right),\\
\tilde{x}_2 &=& y_1 + y_2,\nonumber\\
\tilde{x}_3 &=& \tilde{y}_2 - \tilde{y}_1,\nonumber
\ee
we find
\be\label{liformVI}
  g^{-1}dg &=& (dy_1 \cosh y_3 + dy_2 \sinh y_3) U_1 + 
\nonumber\\
&&(dy_1 \sinh y_3 +
dy_2 \cosh y_3) U_2 + dy_3 U_3.
\ee
Calculating the matrices (\ref{matrices}) for this decomposition, we get
\be
\mu &=& 0,\nonumber\\
\nu &=&
\left(\begin{array}{ccc}
\cosh y_3 & \sinh y_3 & 0\\
\sinh y_3 & \cosh y_3 & 0\\
-(y_2 \cosh y_3 + y_1 \sinh y_3) &
-(y_1 \cosh y_3 + y_2 \sinh y_3) & 1
\end{array}\right),\nonumber\\
M &=& 
\left(\begin{array}{ccc}
0 & -1 & -\frac{t^2}{2} \\
0 & -1 & \frac{t^2}{2}\\
-t^2 & 0 & 0
\end{array}\right),\\
N &=& 
\left(\begin{array}{ccc}
0 & -\frac{t^2}{2} & -1 \\
0 & -\frac{t^2}{2} & 1\\
-1 & 0 & 0
\end{array}\right),\nonumber
\ee
giving us the background
\be
 E &=& 
\left(\begin{array}{ccc}
\frac{1}{t^2}+\frac{t^2}{4} & \frac{1}{t^2}-\frac{t^2}{4}  & 0 \\
\frac{1}{t^2}-\frac{t^2}{4} & \frac{1}{t^2}+\frac{t^2}{4} & 0\\
0 & 0 & t^2
\end{array}\right),\nonumber\\
\phi &=& \phi^{(0)} + \ln t^2.
\ee
To go to a coordinate basis we have to use the left invariant one
forms (\ref{liformVI}) to write the metric as
\be
ds^2 = -dt^2 + \left((\frac{1}{t^2}+\frac{t^2}{4})\cosh 2 y_3
+ (\frac{1}{t^2}-\frac{t^2}{4})\sinh 2y_3\right)\left(dy_1^2 +
dy_2^2\right)\nonumber\\
+ 2\left((\frac{1}{t^2}+\frac{t^2}{4})\sinh 2 y_3
+ (\frac{1}{t^2}-\frac{t^2}{4})\cosh 2y_3\right)dy_1dy_2
+t^2 dy_3^2.
\ee
Since the matrix $E$ is symmetric there is no antisymmetric tensor and
to evaluate the total dilaton contribution we have to evaluate
$\phi^{(0)}$ in the new coordinates. From (\ref{cchange}) we find that
$\phi^{(0)} = 2x_1 = -2y_3$ which gives the final result 
\be
\phi = \ln t^2 - 2 y_3.
\ee
Since $\det E = t^2$ and $\det M = t^4$ in this case, the measure
factor is
\be
 {\cal D} g \frac{t^3 t^2}{t^4} = {\cal D}g \sqrt{\det G},
\ee
where again $G$ is the symmetric part of $E$ for this model so that
the full measure is again the correct coordinate invariant measure of
the sigma model.

\bigskip
\begin{flushleft}
\underline{$(5|2.i)$:}
\end{flushleft}
\bigskip
For the decomposition $(5|2.i)$ we have the decomposition matrix
\be
\left(\begin{array}{c} T\\ \tilde{T} \end{array}\right) =
\left(\begin{array}{cccccc}
-1 & 0 & 0 & 0 & 0 & 0\\
0 & 0 & \frac{1}{2} & 0 & 1 & 0\\
0 & -\frac{1}{2} & 0 & 0 & 0 & 1\\
0 & 0 & 0 & -1 & 0 & 0\\
0 & 1 & 0 & 0 & 0 & 0\\
0 & 0 & 1 & 0 & 0 & 0
\end{array}\right)
\left(\begin{array}{c} U\\ \tilde{U} \end{array}\right),
\ee
and the decomposition of a general group element we choose as
\be
 l = \tilde{g}g = e^{\tilde{y}_3\tilde{U}^3}
 e^{\tilde{y}_2\tilde{U}^2}e^{\tilde{y}_1\tilde{U}^1}
     e^{y_3 U_3} e^{y_2 U_2} e^{y_1 U_1},
\ee
which is equivalent to a change of coordinates $x_1 = -y_1,\;x_2 =
\ldots$\footnote{We only need the expression for $x_1$ in terms of the
new coordinates in this example.}. Since the isometry group and the
choice of coordinates is the 
same as in the first case (Bianchi V) the left invariant form is the
same (\ref{liformV}). The matrices (\ref{matrices}) turn out to be
\be
\mu &=& \left(\begin{array}{ccc}
0 &0 & 0\\
0 & 0 & \sinh y_1\\
0 & -\sinh y_1 & 0
\end{array}\right),\nonumber\\
\nu &=& \left(\begin{array}{ccc}
1 & -y_2 e^{y_1} & -y_3 e^{y_1}\\
0 & e^{y_1} & 0\\
0 & 0 & e^{y_1}
\end{array}\right),\nonumber\\
M &=& \left(\begin{array}{ccc}
-t^2 &0 & 0\\
0 & -1 & 0\\
0 & 0 & -1
\end{array}\right),\\
N &=& \left(\begin{array}{ccc}
-1 &0 & 0\\
0 & -t^2 & -\frac{1}{2}\\
0 & \frac{1}{2} & -t^2
\end{array}\right),\nonumber
\ee
leading to a background
\be
E &=& \left(\begin{array}{ccc}
t^2 & 0 & 0\\
0 & \frac{4t^2}{4t^4+e^{4y_1}} & \frac{-2e^{2y_1}}{4t^4+e^{4y_1}}\\
0 & \frac{2e^{2y_1}}{4t^4+e^{4y_1}} & \frac{4t^2}{4t^4+e^{4y_1}}
\end{array}\right),\nonumber\\
\phi &=& \phi^{(0)} + \ln \left(4t^4 + e^{4y_1}\right) -2y_1.
\ee
The symmetric part of the matrix $E$ gives the metric in the
coordinate basis
\be\label{full}
 ds^2 = -dt^2 + t^2 d y_1^2 + \frac{4t^2e^{2y_1}}{4t^4+e^{4y_1}}
\left(dy_2^2 + dy_3^2\right),
\ee
whereas the antisymmetric part of the matrix $E$ gives the
antisymmetric tensor
\be
 b = -\frac{e^{2y_1}}{2t^2} e^2 \wedge e^3,
\ee
where now the $e$'s refer to the vierbeins of the full metric
(\ref{full}). Finally we get the dilaton by remembering that
$\phi^{(0)}= 2x_1 = -2y_1$ which gives the final result
\be
\phi = \ln\left(4t^4+e^{4y_1}\right)-4y_1.
\ee
Using the relevant expressions the measure factor becomes
\be
{\cal D}g \frac{t^3\frac{4t^2}{4t^4+e^{4y_1}}}{t^2} = {\cal D}g
\sqrt{\det{G}},
\ee
where again $G$ is the symmetric part of $E$.

\bigskip
\noindent
These backgrounds have all been checked to satisfy the string beta
function equations
\be
 0 &=& R_{ab} - \nabla_{a}\nabla_b \phi 
    - \frac{1}{4}H_{acd}H_{b}^{\;cd},\nonumber\\
 0 &=& \nabla^{c}\phi H_{cab} + \nabla^{c}H_{cab},\\
 0 &=& R - 2 \nabla^2 \phi - \nabla_a \phi \nabla^{a}\phi
  -\frac{1}{12}H_{abc}H^{abc}.\nonumber
\ee

\bigskip
\begin{flushleft}
\underline{$(1|6_0)$ and $(5.ii|6_0)$:}
\end{flushleft}
\bigskip
The decompositions $(1|6_0)$ and $(5.ii|6_0)$ are problematic for one and the
same reason. It turns out that the coordinate transformation that is
required to go from the original $(5|1)$ to any of these models is
always of the form $x_1 = f(y,\tilde{y})$ so that the dilaton
$\phi^{(0)}=2x_1$ always depends also on the auxiliary coordinates
$\tilde{y}$ which are to be integrated over to get a sigma model
depending only on $y$. This phenomenon is particularly interesting for
the $(1|6_0)$ model since it just corresponds to ordinary non-abelian
dual of the background $(6_0|1)$ which we just found. The ordinary
non-abelian T-dualization procedure ensures that the dual background
is conformal in the case where the original dilaton depends on
spectators only. In this case, where the dilaton also depends on the
coordinates, it is not clear how the procedure should work. For
instance, how should one gauge the isometry? It is an open question if
the duality can be made to work also in this case.

\section{Discussion}\label{disc}
We have extended the path integral formulation of Poisson-Lie
T-duality to the case of Drinfeld doubles which have more than one
decomposition into bi-algebras. By carefully considering and regulating
determinants which appear in the process of integrating out various
degrees of freedom we have been able to determine which cases gives us
conformal dual sigma models and which not. In particular we were able
to find the correct dilaton shift correcting a mistake in
\cite{us}. As an explicit example we showed that the following
space-times are dual to each other
\begin{flushleft}
\underline{$(5|1)$}
\end{flushleft}
\be
 ds^2 &=& -dt^2 + t^2\left(dx^2 + e^{2x}(dy^2+dz^2)\right),\nonumber\\
 b &=& 0,\\
 \phi &=& 0,\nonumber
\ee
\begin{flushleft}
\underline{$(6_0|1)$}
\end{flushleft}
\be
ds^2 &=& -dt^2 + \left((\frac{1}{t^2}+\frac{t^2}{4})\cosh 2 z
+ (\frac{1}{t^2}-\frac{t^2}{4})\sinh 2z\right)\left(dx^2 +
dy^2\right)\nonumber\\
&&+ 2\left((\frac{1}{t^2}+\frac{t^2}{4})\sinh 2 z
+ (\frac{1}{t^2}-\frac{t^2}{4})\cosh 2z\right)dxdy
+t^2 dz^2,\nonumber\\
b &=& 0,\\
\phi &=& \ln t^2 - 2z,\nonumber
\ee
\begin{flushleft}
\underline{$(5|2.i)$}
\end{flushleft}
\be
 ds^2 &=& -dt^2 + t^2 d x^2 + \frac{4t^2e^{2x}}{4t^4+e^{4x}}
\left(dy^2 + d z^2\right),\nonumber\\
b &=& - \frac{e^{2x}}{2t^2}e^2\wedge e^3,\\
\phi &=& \ln\left(4t^4+e^{4x}\right)-4x.\nonumber
\ee
It is interesting to notice that all three spacetimes have zero scalar
curvature but only the first one is Ricci flat (it is even completely
flat). 

We also showed that the possible duals $(1|6_0)$ and $(5.ii|6_0)$ were
not possible to get in a closed form since the dilaton $\phi^{(0)}$
depends on one of the coordinates which needs to be integrated out in
that case. It would be interesting to understand this better and if it
is possible to anyway go ahead and perform the integration. Presumably
this would introduce some non-trivial dependence on the conformal factor
in the action which could have interesting consequences in
relation with the mixed anomaly terms.

A clear task is to check all duals suggested in section \ref{all} and
check how the duality group acts. The requirement that the duality
group preserves the conformal invariance certainly gives some
restrictions and it would be interesting to see if it is possible to
write down the groups in closed form. It seems that one needs some
constraining principle to get groups of manageable size in general
\cite{moduli}.

Global issues were not addressed at all in this paper. This is a
known problem also in usual non-abelian duality \cite{manda} and we do
not expect it to be any easier to solve here. Only in the abelian case
do we have full control over these effects and can talk about different
descriptions of the same theory. Non-abelian dualities should perhaps
rather be viewed as solution generating techniques, producing one
string background from another.

\acknowledgments{The author is grateful to Libor \v{S}nobl for several
interesting conversations and to Ulf Lindst\"{o}m and Martin Ro\v{c}ek
for useful comments and correspondence. This work was supported by the
Czech Ministry of Education under Contract No. 143100006}

\end{document}